\documentclass[onecolumn,prd]{revtex4}
\usepackage{graphicx}
\begin{document}
%\caption{\label{label}A planar and a volume detector}
%\caption{\label{label}Drawing of the Earth showing a neutrino entering the Earth at a nadir angle $\theta $.}
\title{Event rates vs. cross sections at neutrino telescopes}

\author{Shahid Hussain}

\affiliation{Department of Physcis and Astronomy, University of Delaware, Newark, DE 19716, USA}

\email{shahid@bartol.udel.edu}

\begin{abstract}
One of the major goals of neutrino astronomy is to explore the otherwise
unknown fluxes and interactions of ultrahigh energy neutrinos. The existing
neutrino telescopes look at three types of events: particle showers, muons,
and taus. In this paper we discuss the dependence of the event rates on the
neutrino nucleon cross-sections as we scale the cross sections, with energy,
in different scenarios beyond the standard model. Our focus will be on the
IceCube detector.
\end{abstract}
\maketitle

\textit{Introduction: }Neutrino-nucleon cross sections at ultrahigh energies
are unknown at present. These cross sections might deviate from their
standard model (SM) values due to the turn on of any new-physics processes
above a TeV center of mass energies. A number of neutrino telescopes are
already taking data and others are under developmental stage. One of the
major goals of these telescopes is to determine neutrino-nucleon cross
sections above a PeV. As the neutrino fluxes are not known either, the cross
section information is lost if one merely looks at the absolute event rates
integrated over neutrino energy. To determine the cross sections one needs
to calculate the ratios of the upward and downward event rates. The
knowledge of the scaling of event rates with cross sections also helps
differentiate among the cross section models. In this work we study the
scaling of neutrino event rates with the cross sections. For details see
Ref. \cite{hmms1}. A number of authors have done similar studies \cite
{kw,ffwy,afgs1,afg,icehep}. 

\textit{A simple analytic model: }Here we give a very simple model for the
dependence of neutrino event rates on the neutrino-nucleon cross sections.
We will discuss the case for two generic detectors: an underground volume
detector, like IceCube \cite{ice04} and a planar detector placed
horizontally on the surface of the Earth. The cross section dependence of
the \textit{downward} event rates is trivial. The \textit{downward} shower
rates scale as $\frac{d\Gamma }{dE_{\nu }}\sim \sigma _{t}$, where $\sigma
_{t}$ is the total cross section due to all the processes that produce
showers. Similarly, the lepton event rates are given by $\frac{d\Gamma }{%
dE_{\nu }}\sim \sigma _{cc}$, where $\sigma _{cc}$ is the charged current
(CC) interaction cross-section.

The dependence of the \textit{upward} rates on the cross sections is derived
as follows: \textbf{i)} the rates depend on the upward flux at the detector: 
$\frac{d\Gamma }{dE_{\nu }}\sim F_{d}$, where $F_{d}$ depends on the
attenuation cross section. If one ignores both the down-scattered neutrinos
and the coupling among different neutrino flavors due to tau decays, the
flux at the detector at nadir angle $\theta $ is $F_{d}=F_{0}e^{-l/\lambda
_{a}}$. Here $F_{0}$ is the flux falling on surface of the Earth; $l$ is the
length neutrino travels through the Earth to reach the detector, and $%
\lambda _{a}=1/N_{A}\rho \sigma _{a}$ ($N_{A}$ is Avogadro's number and $%
\rho $ is the density of the Earth which is assumed to be constant in our
simple model). $\sigma _{a}$ is the attenuation cross section defined by $%
\sigma _{a}=\sum_{i}y_{i}\sigma _{i}$. The sum is over different types of
interactions that a neutrino undergoes during its propagation through the
Earth; here $y$ is the average inelasticity of an interaction defined as: $%
y=1$ for a $\nu _{e}$ or $\nu _{\mu }$ charged current interaction; for $\nu
_{\tau }$ charged current interactions, although a neutrino is completely
absorbed, it is regenerated by tau decays; hence the effective $y$ for $\nu
_{\tau }$'s is smaller than 1. For neutral current interactions $y$ is just
the conventional average inelasticity of the interaction ($y\simeq 0.2-0.25$ 
\cite{gqrs}). Given these considerations, for standard model, $\sigma
_{a}(SM)\cong 0.7\sigma _{t}(SM)$ around 10 PeV \cite{hmms1}.\textbf{\ ii)}
The rates depend on the interaction cross section: $\frac{d\Gamma }{dE_{\nu }%
}\sim \sigma _{i}$; for showers, $\sigma _{i}$ is the total cross section
due to the processes that produce showers and for leptons it is the charged
current interaction cross section. \textbf{iii)} The rates depend on the
area projected perpendicular to the direction of incoming neutrino: $\frac{%
d\Gamma }{dE_{\nu }}\sim A_{p}$. For a volume detector, $A_{p}$ is almost a
constant; for a surface detector, $A_{p}$ is a function of the nadir angle;
for example, for a planar detector sitting horizontally on the surface of
Earth, the projected area is $A_{p}=Acos\theta $, where $A$ is the surface
area of the detector. This difference in $A_{p}$ for a volume detector and a
planar detector causes the rates, integrated over angle, to be different for
the detectors as we discuss below. 

The above discussion implies $\frac{d\Gamma }{dE_{\nu }}\sim \frac{1}{%
\lambda _{i}}\int d\Omega A_{p}F_{d}\sim \frac{F_{0}}{\lambda _{i}}%
\int\limits_{0}^{\pi /2}$ $d\theta sin\theta A_{p}e^{-l/\lambda _{a}}$ which
for a volume detector, placed on the surface (depth $d=0$) in our simple
model, reduces to $\frac{d\Gamma }{dE_{\nu }}\sim \frac{\lambda _{a}}{%
\lambda _{i}}(1-e^{-2R/\lambda _{a}})$, and for a planar detector placed
horizontally on the surface of Earth it reduces to $\frac{d\Gamma }{dE_{\nu }%
}\sim \frac{\lambda _{a}^{2}}{\lambda _{i}}(1-e^{-2R/\lambda _{a}}-\frac{2R}{%
\lambda _{a}}e^{-2R/\lambda _{a}})$. Now we can write the scaling
expressions for the event rates in two different limits: For $\lambda
_{i}\ll 2R$, we get: $\frac{d\Gamma }{dE_{\nu }}\sim \frac{\lambda _{a}}{%
\lambda _{i}}=\sigma _{i}/\sigma _{a}$ for a volume detector and $\frac{%
d\Gamma }{dE_{\nu }}\sim \frac{\lambda _{a}^{2}}{\lambda _{i}}=\sigma
_{i}/\sigma _{a}^{2}$ for the planar detector. For $\lambda _{i}\gg 2R$, the
Earth becomes transparent and both for volume and planar detectors we get $%
\frac{d\Gamma }{dE_{\nu }}\sim \sigma _{i}$ (same as downward event rates as
expected).

\textit{Detailed Simulation: }We solve coupled Boltzmann transport equations
for three neutrino flavors and the taus \cite{hm}. As an example, we use
IceCube as our detector and use the WB model flux $F_{0}=6\times
10^{-8}(GeV.cm^{-2}.s^{-1})/(E(GeV))^{2}$ \cite{wb}. We include tau decay
and the down scattering terms in the Boltzmann transport equations. We count
all showers produced within the detector volume and take the energy losses
of muons and taus to be independent of their energies; we also ignore
tagging of taus here; our arguments about the scaling of event rates are not
sensitive to these details. We do the simulation assuming three cases for
neutrino-nucleon cross section models as explained below. Figures 1 and 2
show the results obtained from detailed simulation. For details see Ref. 
\cite{hmms1}.

\begin{figure}[h]
\begin{center}
\includegraphics[width=32pc]{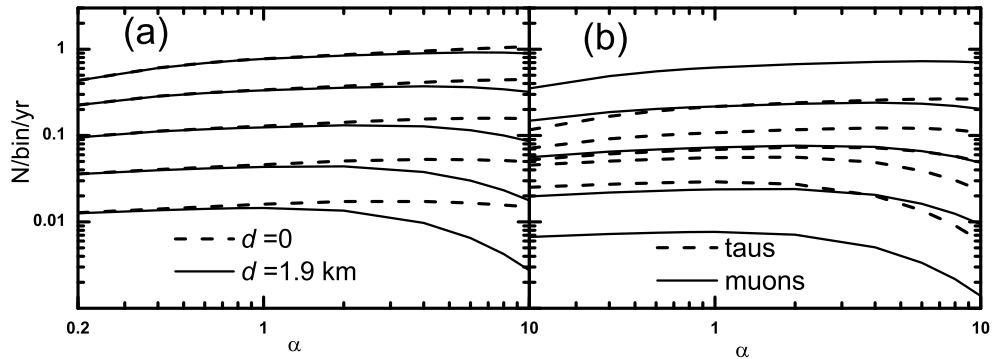}
\caption{\label{label}Expected event rates in IceCube vs. $\alpha $ (see text). Pairs of
curves, from top to bottom, are for $10^{6.5}\leq E_{\nu }\leq 10^{9}$ GeV
in five bins of width $\Delta \log _{10}E_{\nu }=0.5$. Fig 1(a)
shower rates while the Fig. 1(b) shows lepton rates with d=1.9 km.}
\end{center}
\end{figure}

Fig. 1 shows the results for the case where we assume the following
cross-section models: $\sigma _{cc}=\alpha \sigma _{cc}(SM)$ and $\sigma
_{nc}=\alpha \sigma _{NC}(SM)$ (these models are motivated by the fact that
the QCD saturation effects might alter the scaling of neutrino-nucleon cross
sections with energy at ultrahigh energies \cite{glr}). For this case, $%
\sigma _{a}=(1+\alpha )\sigma _{a}(SM)$ and $\sigma _{i}$ [showers, leptons] 
$=$ [$(1+\alpha )\sigma _{t}(SM)$, $(1+\alpha )\sigma _{cc}(SM)$]. Therefore, 
our simple analytic model implies that the rates $\frac{d\Gamma }{dE_{\nu }}\sim \sigma _{i}/\sigma _{a}$ 
are independent of $\alpha $ for these cross section models. Let us compare
this prediction with the detailed simulation shown in Fig. 1. Fig. 1(a)
shows results for the shower rates for a detector with IceCube
characteristics. To illustrate the importance of detector depth $d$, we show
results for $d=1.9$ km (IceCube depth) and $d=0$ in Fig. 1(a). Fig. 1(b)
shows lepton rates for $d=1.9$ km case. We see in Fig.1(a): \textbf{a)} For
higher cross sections and energies, even a depth of 1.9 km makes a
significant difference in the event rates as compared to the $d=0$ case.
This is explained as follows: For $d=1.9$ km case, even for horizontal
neutrinos the distance traveled through the Earth is $l\simeq 220$ km; hence
the rate $\frac{d\Gamma }{dE_{\nu }}\sim \sigma _{i}e^{-220(km)/\lambda _{a}}
$ is an upper limit on how fast the event rates can grow with the cross
section for $d=1.9$ km case. At higher energies and cross sections, $\lambda
_{a}<220(km)$ (e.g. $\sigma (new)=10\sigma (SM)$, $E_{\nu }=10^{9}$ GeV $%
\Rightarrow $ $\lambda _{a}\simeq 50$ km), hence the above upper limit on
the rate is a decreasing function of $\alpha $ for our cross section models.
On the other hand, for $d=0$ case, $\frac{d\Gamma }{dE_{\nu }}\sim \sigma
_{i}/\sigma _{a}$ from our analytic model. These trends are obvious in Fig.
1(a). \textbf{b)} Another extreme limit is the one at low energies and cross
sections (upper left corner in Fig. 1(a)) where both cases, $d=0$ and $d=1.9$
km, are indistinguishable. This is because, in this limit, the Earth is
becoming transparent to the neutrinos. \textbf{c)} For the case between the `%
\textbf{a'} and `\textbf{b'} extreme cases, our analytic model gives $\frac{%
d\Gamma }{dE_{\nu }}\sim \sigma _{i}/\sigma _{a}$; however, in Fig. 1, we
see there is a slight increase in the rates with $\alpha $ which is due to
the details of the simulation that are absent from our analytic model (see
the discussion of Fig. 2 below).

\begin{figure}[h]
\includegraphics[width=18pc]{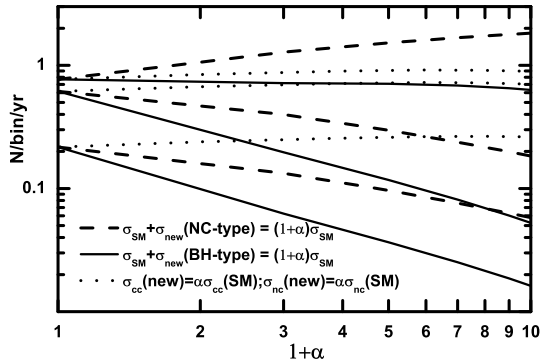}\hspace{2pc}%
\begin{minipage}[b]{18pc}\caption{\label{label}Comparison of rates, for the bin $10^{6.5}\leq E_{\nu }\leq 10^{7}$
GeV, for different cross section models. There are three triplets of curves:
Top: showers. Middle: muons. Bottom: taus.}
\end{minipage}
\end{figure}

Fig. 2 compares the scaling of event rates with the cross sections, for the
energy bin $10^{6.5}\leq E_{\nu }\leq 10^{7}$ GeV, for different types of
cross section models. The purpose of this figure is to illustrate the
importance of inelasticity of the cross sections in the scaling arguments.
All the results are for IceCube with $d=1.9$ km. There are three triplets of
curves: upper, middle, and lower; these triplets correspond to shower, muon,
and tau rates, respectively. Each triplet has three curves (dashed, solid,
and dotted) corresponding to three classes of models (labeled in the legend
in Fig. 2) for the cross sections. We use the following cross section models
in Fig. 2: For the dotted curves, our cross section model is the same as in
Fig. 1. For the solid curves (motivation is black hole formation in low
scale gravity models \cite{add}), in addition to the SM neutral current and
charged current cross sections, we introduce a new cross section with
inelasticity = 1 and we normalize it to the SM total cross section at $%
\alpha =1$. This implies $\sigma _{a}=\sigma _{a}(SM)+\alpha \sigma _{t}(SM)$
and $\sigma _{i}$ [showers, leptons] $=$ [$(1+\alpha )\sigma _{t}(SM)$, $%
\sigma _{cc}(SM)$] $\Rightarrow $ $\sigma _{i}/\sigma _{a}$ [showers,
leptons] $\cong $ [$(1+\alpha )/(0.67+\alpha )$, $0.7/(0.67+\alpha )$]. For
the dashed curves (motivation is graviton exchange in low scale gravity
models.), in addition to the SM neutral current and charged current cross
sections, we introduce a cross section equivalent to the SM neutral current
cross section and we normalize it to the SM\textit{\ total} cross section at 
$\alpha =1$. This implies $\sigma _{a}=\sigma _{a}(SM)+\alpha 0.25\sigma
_{t}(SM)$ and $\sigma _{i}$ [showers, leptons] $=$ [$(1+\alpha )\sigma
_{t}(SM)$, $\sigma _{cc}(SM)$] $\Rightarrow $ $\sigma _{i}/\sigma _{a}$
[showers, leptons] $\cong $ [$(1+\alpha )/(0.67+0.25\alpha )$, $%
0.7/(0.67+0.25\alpha )$]. 

Fig. 2 is in good agreement with our simple analytic model; the event rates
scale like $\sigma _{i}/\sigma _{a}$. The slight deviations from the
expected scaling law ( $\frac{d\Gamma }{dE_{\nu }}\sim \sigma _{i}/\sigma
_{a}$) are due to a couple of details that we have ignored in our simple
model but have included in the simulation (unlike the simulation, our simple
model assumes constant values for Earth density and cross section
inelasticity, and it also ignores the down scattering of neutrinos).

In the above study, the normalization of the event rates was not our
concern. To get an estimate of the event rates in the SM for IceCube, we
have done a more realistic simulation which includes energy dependent energy
losses for muons and taus; for tau energy losses we use approximation III
from Ref. \cite{dhr} and we use energy dependent effective area and
effective volume for IceCube as given in Ref. \cite{icehep}. For WB model,
we get the upward rates per year in the energy bin $10^{6.5}\leq E_{\nu
}\leq 10^{7}$ GeV: $showers=0.7$, $muons=0.8$, and $taus$ $=0.2$ (without
tag); tau rate with tagging (as defined in Ref. \cite{hm} ) is 0.1. The
rates integrated over energy above a PeV are: $showers=2.4$, $muons=2.2$, $%
taus=0.9$ (without tag), and $taus$ with tag $=0.3$. 

\textit{Summary: }We have shown that, in general, the cross section
dependence of the event rates depends on the event type, neutrino energy,
and the cross-section model. \textbf{a) }Downward shower rate is $\frac{%
d\Gamma }{dE_{\nu }}\sim \sigma _{t}$; downward lepton rate is $\frac{%
d\Gamma }{dE_{\nu }}\sim \sigma _{cc}$. \textbf{b)} At neutrino energies
around 1-10 PeV and for a volume detector, within a decade of the change in
cross sections around the SM value, upward shower rate is $\frac{d\Gamma }{%
dE_{\nu }}\sim \sigma _{t}/$ $\sigma _{a}$ and the upward lepton rate is $%
\sim \sigma _{cc}/$ $\sigma _{a}$. For a surface detector, the upward lepton
rate is $\frac{d\Gamma }{dE_{\nu }}\sim \sigma _{cc}/$ $\sigma _{a}^{2}$. 
\textbf{c)} For energies around a PeV and cross sections around 0.1 of the
SM value, the Earth becomes transparent and the upward rates scale with the
cross-section like the downward rates. \textbf{d)} As one goes to higher
energies and/or cross sections, the detector depth becomes important; even a
depth of $d=2$ km can give a significantly different scaling of the rates
with the cross sections, as compared to the $d=0$ case (Fig. 1(a)). \textbf{%
e)} Scaling of the rates with the cross sections is significantly different
for different models of cross sections (Figs. 2).

{\Large Acknowledgments}

This research was supported by the NSF under Grants OPP-0338219 and ANT-0602679. 
I am thankful to D. W. McKay, D. Seckel, and D. Marfatia  for their guidance and support for this research.


\begin{thebibliography}{99}
\bibitem{hmms1}  S. Hussain \textit{et al.}, \textit{preprint}
hep-ph/0606246, to be published in Phys. Rev. Letters.

\bibitem{kw}  A.~Kusenko and T.~Weiler, Phys. Rev. Lett. \textbf{88}, 161101
(2002); %%CITATION = HEP-PH 0106071;%%
S.~Palomares-Ruiz, A.~Irimia and T.~Weiler, Phys. Rev. D \textbf{73}, 083003
(2006).

\bibitem{ffwy}  J.~Feng \textit{et al.}, %P.~Fisher, F.~Wilczek and T.~Yu,
Phys.\ Rev.\ Lett.\ \textbf{88}, 161102 (2002). 
%%CITATION = HEP-PH 0105067;%%

\bibitem{afgs1}  L.~Anchordoqui \textit{et al.}, 
%L.~A.~Anchordoqui, J.~L.~Feng, H.~Goldberg and A.~D.~Shapere,
Phys.\ Rev.\ D \textbf{65}, 124027 (2002). %%CITATION = HEP-PH 0112247;%%

\bibitem{afg}  %\mbox{L.~Anchordoqui {\it et al.},
L.~A.~Anchordoqui, J.~L.~Feng and H.~Goldberg, Phys. Rev. Lett. \textbf{96},
021101 (2006). %%CITATION = HEP-PH 0504228;%%

\bibitem{icehep}  L.~Anchordoqui and F.~Halzen, \textit{preprint }%
hep-ph/0510389. %%CITATION = HEP-PH 0510389;%%

\bibitem{ice04}  J.~Ahrens \textit{et al.}, Astropart.\ Phys.\ \textbf{20},
507 (2004). %%CITATION = ASTRO-PH 0305196;%%

\bibitem{gqrs}  R. Gandhi \textit{et al.}, Astropart. Phys. \textbf{5}, 81
(1996).

\bibitem{hm}  S.~Hussain, D.~McKay, Phys.\ Rev.\ D \textbf{69}, 085004
(2004). %%CITATION = HEP-PH 0310091;%%

\bibitem{wb}  E.~Waxman, J.~Bahcall, Phys.\ Rev.\ D \textbf{59}, 023002
(1999). %%CITATION = HEP-PH 9807282;%%

\bibitem{glr}  L.~Gribov, E.~Levin and M.~Ryskin, Phys.\ Rept.\ \textbf{100}%
, 1 (1983).

\bibitem{add}  N.~Arkani-Hamed, S.~Dimopoulos and G.~R.~Dvali, Phys.\ Lett.\
B \textbf{429}, 263 (1998); %%CITATION = HEP-PH 9803315;%%
Phys.\ Rev.\ D \textbf{59}, 086004 (1999); %%CITATION = HEP-PH 9807344;%%
I.~Antoniadis \textit{et al.}, 
%I.~Antoniadis, N.~Arkani-Hamed, S.~Dimopoulos and G.~R.~Dvali,
Phys.\ Lett.\ B \textbf{436}, 257 (1998). %%CITATION = HEP-PH 9804398;%%

\bibitem{dhr}  S.~Dutta, Y.~Huang and H.~Reno, Phys. Rev. D \textbf{72},
013005 (2005).
\end{thebibliography}
\end{document}